\documentclass[conference]{IEEEtran}

\usepackage{amsmath}

\usepackage[normalem]{ulem}

\usepackage[noadjust, compress]{cite}

\ifCLASSINFOpdf
   \usepackage[pdftex]{graphicx}
\else
\fi

\begin{document}
\title{Estimation of the Soil Water Characteristics \\from Dielectric Relaxation Spectra - a Machine Learning Approach}

\author{\IEEEauthorblockN{Norman Wagner}
\IEEEauthorblockA{Institute of Material Research and Testing\\ at the Bauhaus-University Weimar\\Coudraystr. 4, 99423 Weimar, Germany}
\and
\IEEEauthorblockN{Frank Daschner}
\IEEEauthorblockA{Microwave Group\\ Faculty of Engineering\\ University of Kiel, Germany}
\and
\IEEEauthorblockN{Alexander Scheuermann and Moritz Schwing}
\IEEEauthorblockA{Geotechnical Engineering Centre\\
The University of Queensland\\ Brisbane, Australia}
}

\maketitle

%%-----------------------------------------------------
\begin{abstract}
%\boldmath
The frequency dependence of dielectric material properties of water saturated
and unsaturated porous materials such as soil is not only disturbing in
applications with high frequency electromagnetic (HF-EM) techniques but also
contains valuable information of the material due to strong contributions by
interactions between the aqueous pore solution and mineral phases. Hence,
broadband HF-EM sensor techniques enable the estimation of soil
physico-chemical parameters such as water content, texture, mineralogy, cation
exchange capacity and matric potential. In this context, a multivariate (MV) 
machine learning approach (principal component regression, partial least 
squares regression, artificial neural networks) 
was applied to estimate the Soil Water Characteristic Curve (SWCC)
from experimentally determined dielectric relaxation spectra of a silty clay
soil. The results of the MV-approach were compared with results obtained from
empirical equations and theoretical models as well as a novel
hydraulic/electromagnetic coupling approach. The applied MV-approach gives
evidence, (i) of a physical relationship between soil dielectric relaxation
behavior and soil water characteristics as an important hydraulic material
property and (ii) the applicability of appropriate sensor techniques for the
estimation of physico-chemical parameters of porous media from broadband
measured dielectric spectra.

%In this context, a multivariate approach is applied
%for the simultaneously determination of soil water content,
%porosity and matric potential as hydraulic property
%from measured frequency dependent dielectric material properties of a silty clay soil.

\end{abstract}
% math in the abstract anyway. no keywords

% \fi
\maketitle

%%-----------------------------------------------------
\section{Introduction}

The success of water content estimation in porous media with high frequency
(radio and microwave) non and minimal invasive electromagnetic (HF-EM)
measurement techniques is caused by the dipolar character of the water
molecules resulting in a high permittivity in comparison to other phases. In
particular, HF-EM sensing techniques such as ground penetrating radar (GPR,
\cite{Jol2009, Huis03, Preko2012, Busch2013}), conventional time or frequency
domain domain reflectometry (TDR/FDR, \cite{Robinson2003, Robinson2008,
Schwartz2009}), spatial time domain reflectometry (spatial TDR,
\cite{Scheuermann2009a, Rings2010, Huisman2010, Vereecken2013}), capacitance
methods \cite{Evett2009} as well as active and passive microwave remote sensing
techniques (\cite{ Kerr2012, Bronstert2012}) offer the possibility to monitor
spatial and temporal variations of soil physical state parameters, e.g. the
volume fraction of free pore water, with high resolution (\cite{Huis03,
Robinson2008, Scheuermann2009a, Huisman2010, Preko2012}). Hence, the objective
of numerous experimental and theoretical investigations are the development of
generalized electromagnetic models for a broad class of soil textures and
structures (\cite{Topp80, Roth1990, Miyamoto2005}). Mostly, these empirical, 
numerical or theoretical approaches are based on the assumption of a constant
dielectric permittivity of the soil as a function of volumetric water content
in a narrow frequency and temperature range (\cite{Sihv00, Evet05, Rega06,
Mironov2004}). However, HF-EM techniques cover a broad frequency range between
approximately 10~MHz, in the case of spatial TDR, to at least 10~GHz in X-band
remote sensing applications (\cite{Beha05, Robinson2008}). For this reason, the
knowledge of the frequency and temperature dependent HF-EM material parameters
is needed for a successful application and combination of the different sensing
techniques. However, interactions between the aqueous pore solution and solid
phases lead to strong contributions to the electromagnetic material properties
due to interphase processes (\cite{WagScheu2009a, Wagner2011}). Therefore, the
broadband dielectric spectrum contains valuable information about porous media
and it can be used for an estimation of physico-chemical parameters besides
free pore water such as texture, structure, mineralogy, cation exchange
capacity and matric potential as important hydraulic property with broadband
HF-EM sensor techniques. In this context, a multivariate (MV) approach
according to Daschner et al. (2003) \cite{Daschner2003} was applied for the
simultaneously determination of soil water content, porosity and matric
potential from measured frequency dependent dielectric material properties. The
results of the MV-approach were compared with results obtained from empirical
equations and theoretical models as well as a novel hydraulic/electromagnetic
coupling approach.

%-----------------------------------------------------%-----------------------------------------------------
\section{Modeling of soil dielectric relaxation behavior}\label{sec:model}
%\section{\normalsize{COUPLING HYDRAULIC AND DIELECTRIC SOIL PROPERTIES}}
%-----------------------------------------------------%-----------------------------------------------------

Organic free soil as a typical porous material mainly consists of four phases:
solid particles (various mineral phases), pore air, pore fluid as well as a
solid particle - pore fluid interface. In principle the fractions of the soil
phases vary both in space (due to composition and density of the soil) and time
(due to changes of water content, porosity, pore water chemistry and
temperature). The electromagnetic properties of the solid particles are
frequency independent in the considered temperature-pressure-frequency range.
Real relative permittivity varies from 3 to 15 \cite{Robinson2004}. The pore
fluid as well as interface fluid are mainly an aqueous solution with a
temperature-pressure-frequency dependent relative complex permittivity
$\varepsilon^\star_{w} (\omega ,T, p)$ according to the modified Debye model
\cite{Kaat07}:
%-----------------------------------------------------
\begin{equation}\label{eq:Debye}
\varepsilon^\star_{w} (\omega ,T,p) - \varepsilon _\infty(T,p)   =
\frac{{\Delta \varepsilon (T,p) }}{1 + j\omega
\tau_w(T,p)}-j\frac{\sigma_{w}(T,p)}{\varepsilon_0\omega},
\end{equation}
%-----------------------------------------------------
with direct current conductivity $\sigma_{w}$, high frequency limit of
permittivity $\varepsilon_\infty$ and relaxation strength
$\Delta\varepsilon=\varepsilon_S+\varepsilon_\infty$ with static dielectric
permittivity $\varepsilon_S$. Under atmospheric conditions the dielectric
relaxation time of water $\tau_w(T)$ depends on temperature $T$ according to
the Eyring equation \cite{WagScheu2009a} with Gibbs energy or free enthalpy of
activation $\Delta G_w^\ddag(T) = \Delta H_w^\ddag (T) -T\Delta S_w^\ddag (T)$,
activation enthalpy $\Delta H_w^\ddag(T)$ and activation entropy $\Delta
S_w^\ddag(T)$. Furthermore, Gibbs energy of the interface fluid $\Delta
G_d^\ddag(T)$ is assumed to be a function of the distance from the particle
surface (for quantitative approaches see \cite{WagScheu2009a}). Soil matric
potential $\Psi_m$ is a measure of the bonding forces on water in the soil and
is related to the chemical potential of soil water $\Delta\mu_W = \mu_W^\circ -
\mu_W = \Psi_m V_W$ with chemical potential at a reference state $\mu_W^\circ$
and molar volume of water $V_W$ (\cite{Iwata1995, Hilh01}). Thus, Hilhorst et
al. (\cite{Hilhorst1998, Hilh01, Hilhorst2000}) suggested an approach for the
relationship between $\Psi_m$ and $\Delta G_d^\ddag$:
%-----------------------------------------------------%-----------------------------------------------------
\begin{equation}\label{eq:matric1}
\Psi_m(T)\cdot V_W = \Delta G_{w}^{\ddag\circ}(T) - \Delta G_d^\ddag(T)
\end{equation}
%-----------------------------------------------------%-----------------------------------------------------
with Gibbs energy of water at a reference state $\Delta G_{w}^{\ddag\circ}(T)$
($10.4$\hspace{0.1cm}kJ/mol at atmospheric conditions and $T$=293.15 K). This
relationship is used to calculate Gibbs energy or free enthalpy of dielectric
activation of interfacial water $\Delta G_d^\ddag(T)$. The complex relative
dielectric permittivity of free and interface water of a porous material, e.g.
soil, in dependence of the volumetric water content $\theta$ under atmospheric
conditions then can be calculated \cite{Wagner2013JGR}:
%-----------------------------------------------------%-----------------------------------------------------
\begin{equation}\label{eq:matric2}
 \varepsilon_{a(\theta, n)}^\star (\theta ,T) = \int\limits_{\Psi_m(0)}^{\Psi_m(\theta)}  {\varepsilon^{\star a(\theta, n)}_w(\Psi_m(\theta)
 ,T)}\frac{d\theta(\Psi_m)}{d\Psi_m}d\Psi_m.
\end{equation}
%-----------------------------------------------------%-----------------------------------------------------
The parameter $0\leq a \leq 1$ is defined by the used mixture approach to
obtain the effective complex relative permittivity of the soil
$\varepsilon^\star_{r,\mbox{eff}} (\theta,T )$. The parameter $a$ contains in
principle structural information of free and interface water in the soil and is
strictly speaking a function of volumetric water content $\theta$ and porosity
$n$. However, Hilhorst 1998 (HIL, \cite{Hilhorst2000}) suggests the following
equation
%-----------------------------------------------------%-----------------------------------------------------
\begin{equation}\label{eq:HilMix}
 \varepsilon^\star_{r,\mbox{eff}} (\theta,T ) =
 \frac{\varepsilon_{1}^\star (\theta,T )}{3(2n - \theta )}  + (1 - n)\varepsilon _G(T) + (n - \theta)
\end{equation}
%-----------------------------------------------------%-----------------------------------------------------
with porosity $n$ and real relative permittivity of solid grain $\varepsilon
_G$. As an alternative approach Wagner et al. 2011 \cite{Wagner2011} suggests
the so called advanced Lichtenecker and Rother Model (ALRM):
%-----------------------------------------------------%-----------------------------------------------------
\begin{equation}\label{eq:LRMix}
\varepsilon_{r,\mbox{eff}}^{\star a(\theta, n)} (\theta,T ) =
\varepsilon_{a(\theta, n)}^\star(\theta,T ) + (1 - n)\varepsilon
_G(T)^{a(\theta, n)} + (n - \theta )
\end{equation}
%-----------------------------------------------------%-----------------------------------------------------
with the following empirical relationships for the structure parameter $a$ and
pore water conductivity $\sigma_w^\circ$ at a reference state (T=298.15~K,
\cite{Hayashi2004}) with constants $A_i$, $B_i$, $C_i$:
%---------------------------------------------------------------------------------------------
\begin{equation}\label{eq:LichtRothM2}
 a (\theta ,n) = A_{1}  + B_{1} n^2  + C_{1} \left(\frac{\theta}{n}\right)^2
\end{equation}
%as well as
\begin{equation}\label{eq:LichtRothM3} \log \left( {\sigma^\circ_{w} (\theta
,n)} \right) = \log A_{2}  + B_{2} n + C_{2}  \left(\frac{\theta}{n}\right).
\end{equation}
%---------------------------------------------------------------------------------------------
In the proposed electromagnetic/hydraulic coupling approach the ALRM is used in
complex refractive index model (CRIM) form with a constant structure parameter
$a=0.5$ \cite{WagScheu2009a}.
%and
%\begin{equation}\label{eq:LichtRothM4}
%\sigma_{w}=\sigma^\circ_{w}(1+0.0187(T-298.15)).
%\end{equation}
%\hfill January 11, 2007
%%-----------------------------------------------------
\section{Material and methods}

A slightly plastic clay soil was investigated (for soil details see
\cite{Wagner2011a}). The mineralogical composition is dominated by
tectosilicates (36 wt.\%), carbonates (36~wt\%) and mica (16~wt\%) with a
certain amount of clay minerals (smectite 9~wt.\%, kaolinite 3~wt\%). The
permittivity of the solid particles with 5.57 was estimated from mineralogical
composition (see \cite{Wagner2013JGR} for details).

\subsection{Hydraulic and mechanical soil properties}

\begin{figure}[t]
 \center
  \includegraphics[scale=0.5]{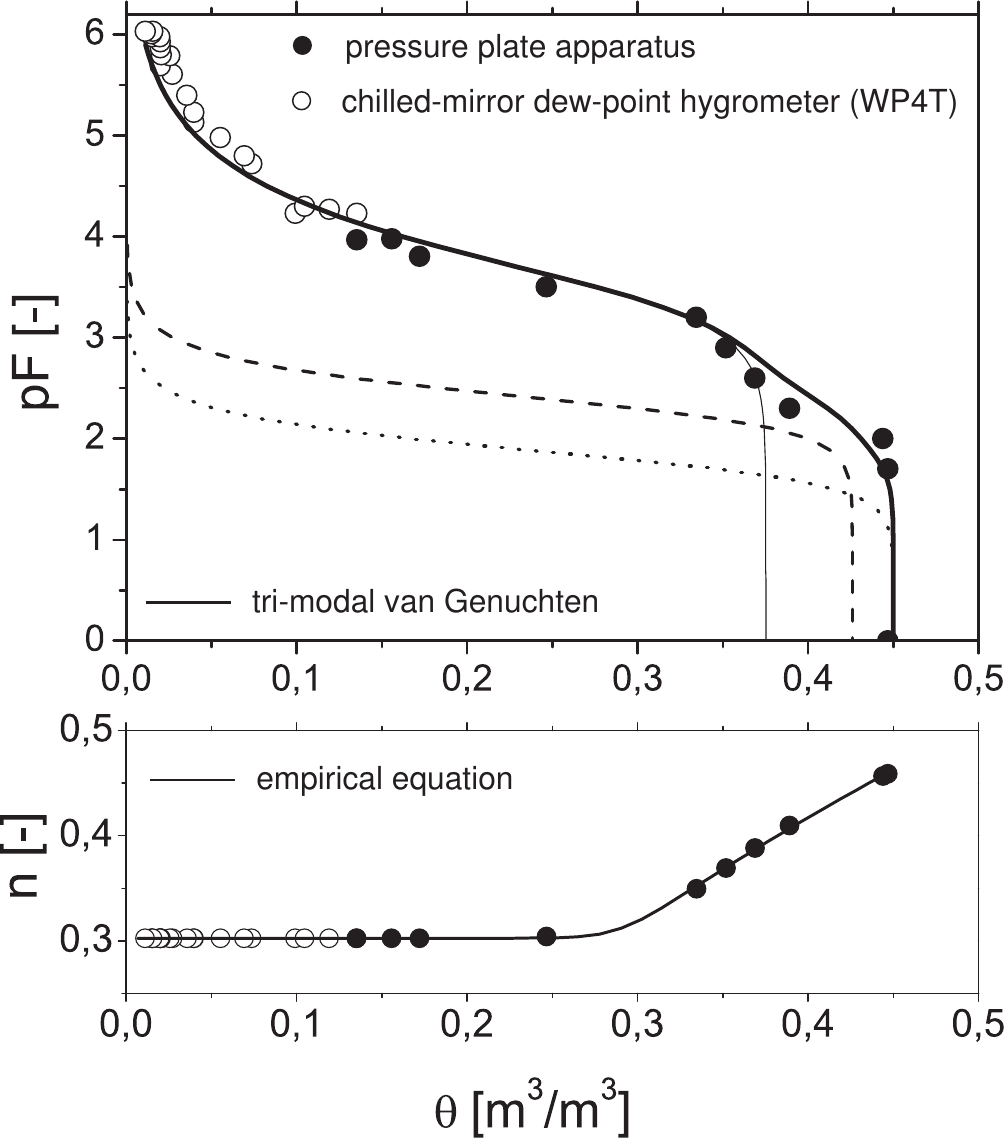}
  \caption{(top) Matric potential $\Psi$ expressed in terms of pF=$\log(|\Psi|
  [hPa])$ and (bottom) porosity $n$ as a function of volumetric water content $\theta$.}\label{fig:Fig2}
\end{figure}

Soil water characteristic curve (SWCC- relationship between volumetric water
content $\theta$ and matric potential $\Psi$) as well as shrinkage behavior
(changes in porosity $n$ as a function of volumetric water content $\theta$)
were determined in separate experimental investigation (for details see
\cite{Schwing2010}, Fig. \ref{fig:Fig2}). SWCC was parameterized with a
tri-modal van Genuchten equation according to Priesack and Durner (2011)
\cite{Priesack2006} using a shuffled complex evolution metropolis algorithm
(SCEM-UA, \cite{Vrugt2003}) and shrinkage behavior was parameterized with an
empirical equation to determine appropriate matric potential and porosity at
defined volumetric water contents (c.f. \cite{Wagner2013JGR}).
%-------------------------------------------------------------
%-------------------------------------------------------------
\subsection{Soil dielectric relaxation spectra} 
\begin{figure}[t]
\center
  \includegraphics[scale=0.56]{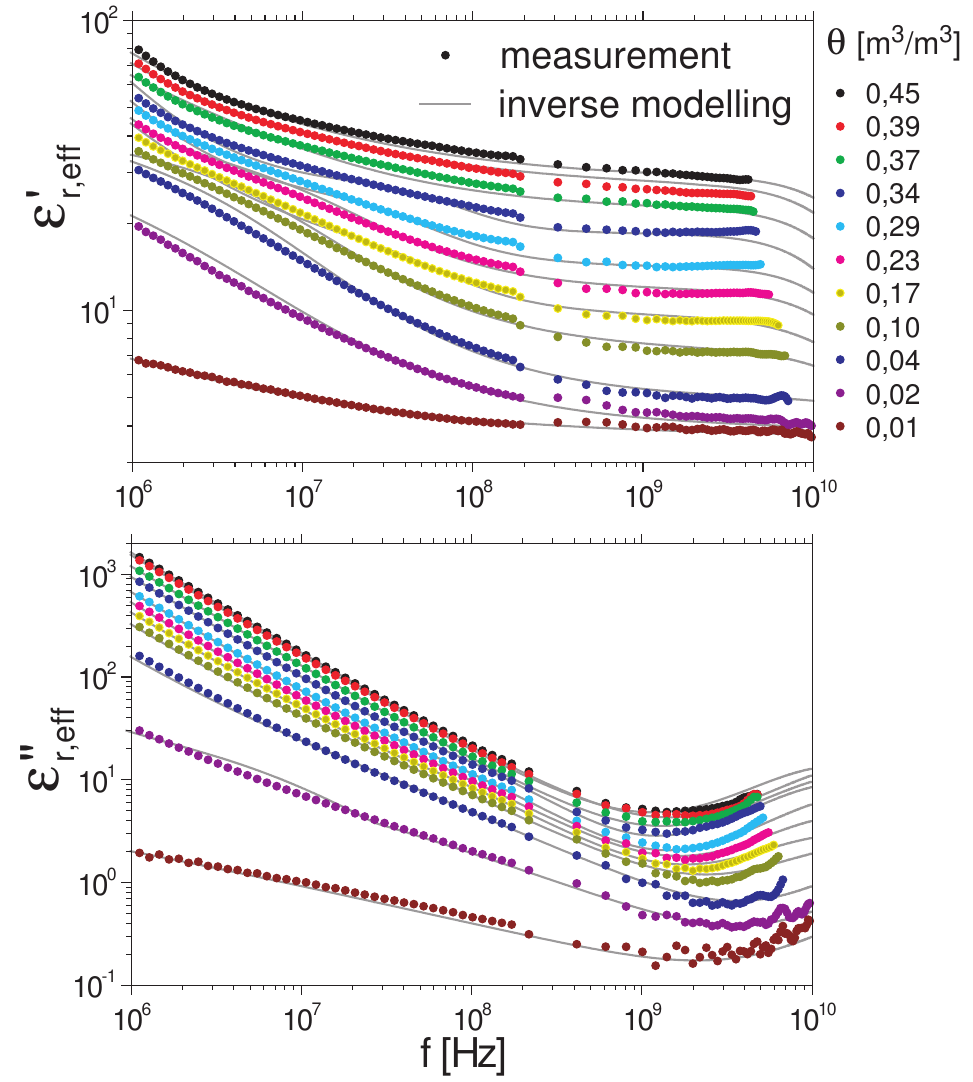}
  \caption{Real part $\varepsilon'_{r,\mbox{eff}}$ and imaginary part $\varepsilon''_{r,\mbox{eff}}$ of the experimental determined complex relative effective permittivity $\varepsilon_{r,\mbox{eff}}^\star$ as a function of frequency for different volumetric water contents $\theta$.}\label{fig:Fig1}
\end{figure}

The frequency dependent complex
permittivity was determined in the frequency range from 1~MHz to 10~GHz with
network analyzer technique according to Wagner et al. (2011) \cite{Wagner2011a}
by means of quasi-analytical methods as well as numerical inversion of measured
four complex S-parameters based on a transverse electrical and magnetical (TEM)
forward model in combination with a broadband electromagnetic transfer function
(Generalized Dielectric Response - GDR, Fig. \ref{fig:Fig1}):
%-------------------------------------------------------------
\begin{equation} \label{eq5}
{\varepsilon }_{\text{r,eff}}^{{\ast }}\left(
{\omega } \right){-}{\varepsilon }_{{\infty }}{=}\sum\limits_{{i=1}}^{N}
\frac{{\Delta }{\varepsilon }_{{i}}}{\left( {j\omega }{\tau }_{{i}} \right)^{{a
}_{{i}}}{+}\left( {j\omega }{\tau }_{{i}} \right)^{{b }_{{i}}}}
{-j}\frac{{\sigma'}_{\text{DC}}}{{\omega }{\varepsilon }_{0}},
\end{equation}
%-------------------------------------------------------------
with ${\varepsilon}_{{\infty}}$ the high-frequency limit of relative
permittivity, ${{\Delta \varepsilon }}_{{i}}$ the relaxation strength, ${\tau
}_{{i}}$ the relaxation time, ${0\le }{a}_{{i}}{,}{b}_{{i}}{\le 1}$ stretching
exponents of the $i$-th process, and ${\sigma'}_{\text{DC}}$ apparent direct
current electrical conductivity.

For the determination of the spectra, the soil sample was in a first step
saturated with deionized water and prepared at liquid limit with gravimetric
water content w~=~0.267~g/g (volumetric water content
$\theta$~=~0.45~m$^3$/m$^3$). The obtained soil suspension was placed in a rod
based transmission line cell (R-TML, see \cite{Wagner2011a}). In a next step
the sample was stepwise dried isothermal at 23~$^\circ$C under atmospheric
conditions at defined humidities and equilibrated. Appropriate mass loss and
sample volume change were estimated during the drying process to obtain
appropriate volumetric water content. At each step a broadband electromagnetic
measurement was performed.
%-----------------------------------------------------
%-----------------------------------------------------

\subsection{Preprocessing of the relaxation spectra}
%-----------------------------------------------------
Prior to the application of the MV-methods the measured dielectric relaxation
spectra were reduced to 81 frequency points in the frequency range from 1~MHz
to 5~GHz. The spectra were compiled in a $m \times 2n$ relative permittivity
matrix
%-----------------------------
    \begin{equation}\label{eq:MLR}
        \uuline{\varepsilon}=
                \left[
                    \begin{matrix}
                        \varepsilon'_{1,1}(f_1) & \cdots & \varepsilon'_{1,n}(f_n) &
                        \varepsilon''_{1,1}(f_1) & \cdots & \varepsilon''_{1,n}(f_n)\\
                         \vdots & & \vdots & \vdots & & \vdots\\
                        \varepsilon'_{m,1}(f_1) & \cdots & \varepsilon'_{m,n}(f_n) &
                        \varepsilon''_{m,1}(f_1) & \cdots & \varepsilon''_{m,n}(f_n)
                    \end{matrix}
                \right]
    \end{equation}
%-------------------------------
for $m$ measurements at $n$ frequencies $f_l$ with real $\varepsilon'_{k,l}$
and imaginary part $\varepsilon''_{k,l}$. The data-set then was transformed in
its principal components $\mathbf{P}$ (PCs)
%-----------------------
    \begin{equation}\label{eq:MLR}
        \mathbf{P}=\mathbf{X}\mathbf{E}
    \end{equation}
%-----------------------
based on a singular value decomposition with loadings (matrix of eigenvectors)
$\mathbf{E}$ and mean-centered and standardized original data matrix
$\mathbf{X}$
%----------------------
\begin{equation}\label{eq:MLR}
        \mathbf{X}=
                \left[
                        \uuline{\varepsilon}-
                        \left[
                            \begin{matrix}
                            1\\\vdots\\1
                            \end{matrix}
                        \right]
                        \left[\overline{\varepsilon'_{1}}\cdots\overline{\varepsilon'_{n}}\hspace{0.2cm}
                              \overline{\varepsilon''_{1}}\cdots\overline{\varepsilon''_{n}}
                        \right]
                \right]
                \left[
                    \begin{matrix}
                        \sigma_{\varepsilon'_{1}}^{-1} & \cdots & 0\\
                        \vdots & \ddots & \vdots\\
                        0 & \cdots & \sigma_{\varepsilon''_{n}}^{-1}
                    \end{matrix}
                \right]
    \end{equation}
%------------------------------
herein $\overline{\varepsilon'_{l}}$ or $\overline{\varepsilon''_{l}}$ denotes
the mean and $\sigma_{\varepsilon'_{l}}$ or $\sigma_{\varepsilon''_{l}}$  the
standard deviation of the data-set at each frequency. Based on principal
component analysis (PCA) an appropriate lower limit of the PCs variance with
$\sigma_{C}=10^{-5}$ was estimated as a robust cut off criterium in the
multivariate calibration step.
%-----------------------------------------------------
\begin{figure*}[t]
  \center
  \includegraphics[scale=0.65]{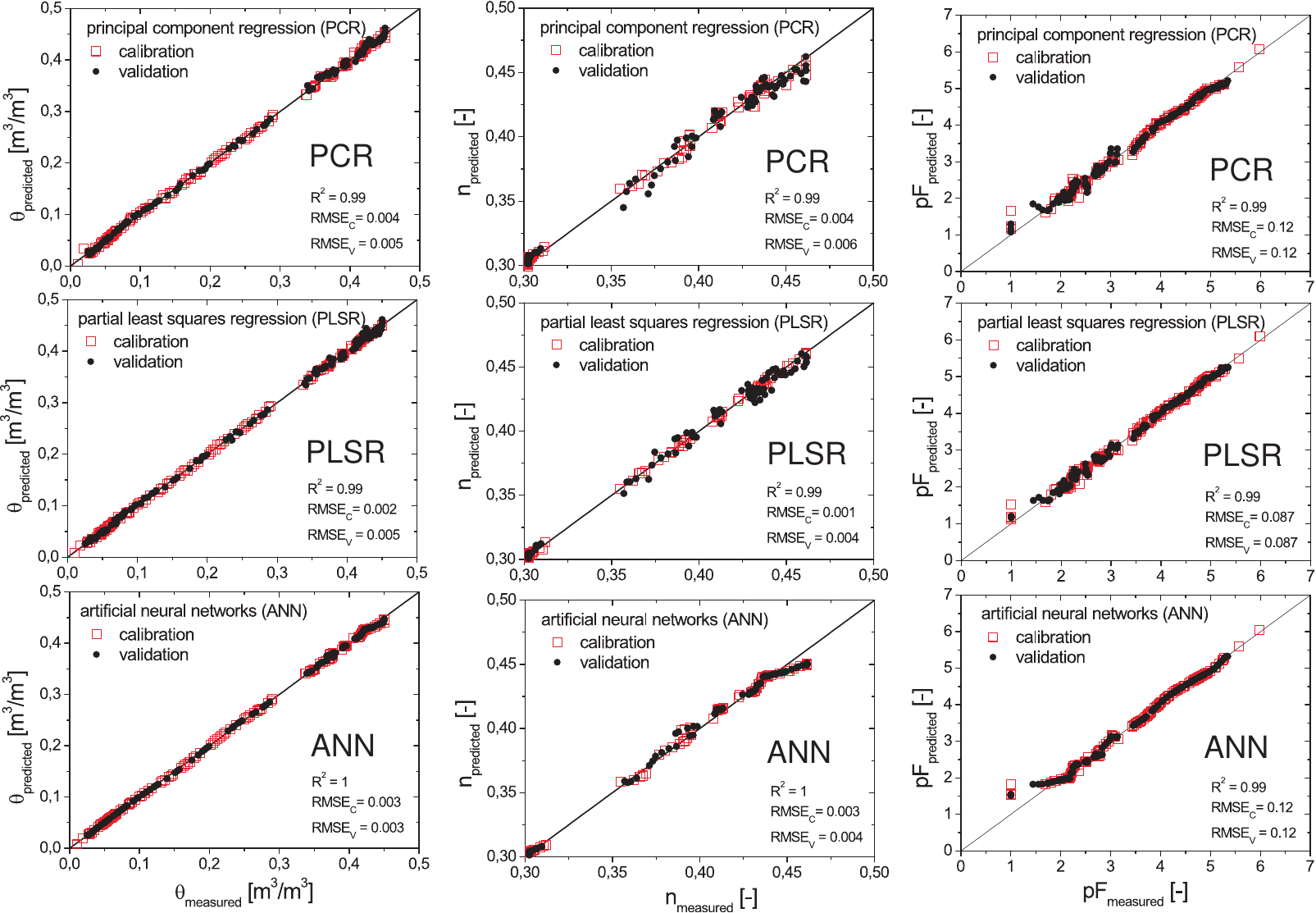}
  \caption{Experimental determined parameters (volumetric water content $\theta$, porosity $n$,  matric potential $\Psi$
  expressed in terms of pF=$\log(|\Psi| [hPa])$) versus predicted results from the dielectric relaxation spectra.}
  \label{fig:Fig3}
\end{figure*}
%-----------------------------------------------------
%-----------------------------------------------------
\begin{figure*}[t]
 \center
  \includegraphics[scale=0.6]{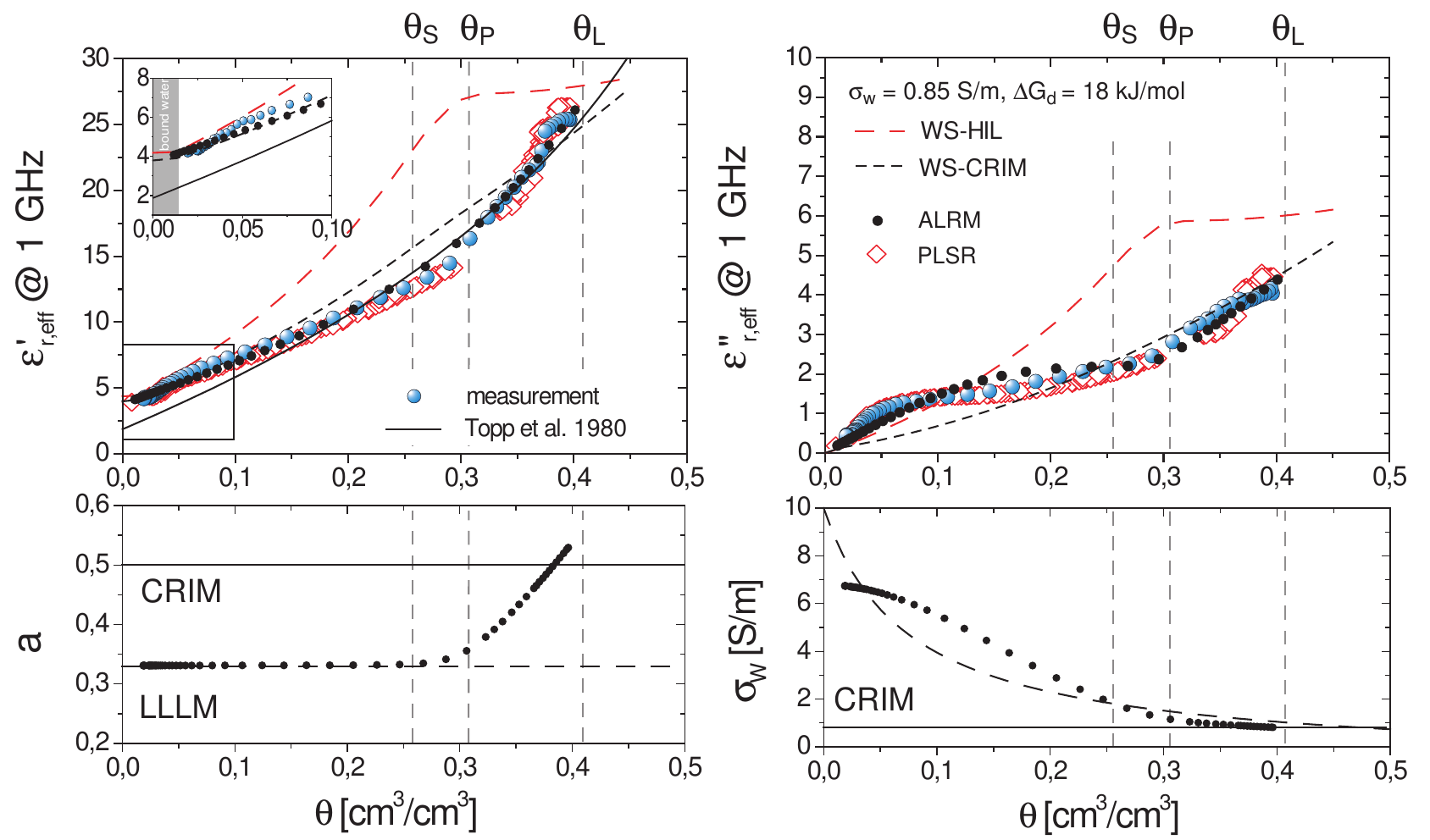}\\
  \caption{{(left, top) Real part $\varepsilon_{r,\mbox{eff}}'$  and (right, top) imaginary part $\varepsilon_{r,\mbox{eff}}''$ of
   complex effective permittivity $\varepsilon_{r,\mbox{eff}}^\star$ at a measurement frequency of 1 GHz as
   a function of volumetric water content $\theta$ in comparison to the empirical calibration function according to Topp et al. 1980 \cite{Topp80},
   WS-HIL, WS-CRIM, ALRM and PLSR (see text for details).
   Volumetric water content at the shrinkage limit $\theta_S$, plastic limit $\theta_P$ and at the liquid
   limit $\theta_L$ are indicated. (left, bottom) Structure parameter $a$ and (right, bottom) effective
    pore water conductivity $\sigma_{W}$ as a function of $\theta$}.}\label{fig:Fig6}
\end{figure*}
%----------------------------------------------------%-------------------------------------------------------

\subsection{Multivariate calibration}

The following multivariate methods were applied to quantitatively relate the
spectra to soil physical properties such as water saturation $S_W$ and porosity
$n$ or volumetric water content $\theta=S_W\cdot n$ as well as matric potential
$\Psi$: principal component analysis with principal component regression (PCR),
partial least squares regression (PLSR). In addition artificial neural networks
(ANN) was applied to the mean-centered and standardized original data matrix
$\mathbf{X}$ using one hidden layer, which contains 10 neurons. The activation
functions of the neurons in the hidden layer are nonlinear (tansig-function)
while those of the output layer are linear (for details see \cite{Daschner2003,
Kent2010}).

%%-----------------------------------------------------
\section{Results}

The complete data-set of 266 measured spectra were randomly divided into two
sets with each 133 groups. One set is used for calibration and one set for
validation. In Table \ref{tab:RMSE} and Figure \ref{fig:Fig3} the results of
the different methods are summarized. The PLSR-technique gives the best results
with lowest RMSEs (see Figure \ref{fig:Fig2} and Tab. \ref{tab:RMSE}).
%---------------------------------------------------------------------------------------------
\begin{table}[t]
\small
  \centering
  \caption{Results of the used multivariate methods (R$^2$=0.99 for all properties and methods) and root
  mean square error estimate for calibration RMSE$_C$ or validation RMSE$_V$, respectively.
      }\label{tab:RMSE}
\begin{tabular}{p{2cm}|lll}
  \hline
            & PCR & PLSR & ANN \\\hline\hline
  $\theta$ [\%]  &  0.4 / 0.5  &  0.2 / 0.5 & 0.3 / 0.3  \\
  $n$ [\%]       &  0.4 / 0.6  &  0.1 / 0.4 &  0.3 / 0.4\\
  $\Psi$ [pF]    &  0.12 / 0.12 &  0.09 / 0.09 &  0.12 / 0.12\\
  \hline
\end{tabular}
\end{table}
%-----------------------------------------------------
The results of the PLSR-approach were compared with the well known
empirical calibration function according to Topp et al. 1980 \cite{Topp80} and
the advanced Lichtenecker and Rother model (ALRM) according to
equation (\ref{eq:LRMix}) to (\ref{eq:LichtRothM3}). Furthermore the
theoretical mixture rule according to equation (\ref{eq:HilMix}) as well as
(\ref{eq:LRMix}) in four phase CRIM form were used considering soil water
characteristic curve (SWCC) as well as shrinkage behavior (Fig. \ref{fig:Fig2})
based on the improvements suggested in section \ref{sec:model} (WS-HIL,
WS-CRIM) to calculate complex relative permittivity of free and interface water
according to (\ref{eq:matric2}).

In Fig. \ref{fig:Fig6} theoretically calculated, statistically estimated and
experimentally determined $\varepsilon_{r,\mbox{eff}}^\star$ at 1 GHz is
represented for WS-CRIM, WS-HIL, ALRM, PLSR and Topp. The models (WS-HIL,
WS-CRIM, ALRM) predict the permittivity at very low water content whereas the
Topp-equation clearly underestimate $\varepsilon'_{r,\mbox{eff}}$. The
frequency and water content dependent complex effective relative permittivity
is poorly predicted with WS-HIL while the qualitative characteristics is
similar. Substantially better results gives WS-CRIM. The deviation between
WS-CRIM and experimentally obtained imaginary part of effective complex
permittivity $\varepsilon_{r,\mbox{eff}}^\star$ especially below the shrinkage
limit $\theta_S$ indicates the dependence of the so called structure exponent
as well as pore water conductivity on volumetric water content or porosity and
saturation pointed out by Wagner et al. 2011 \cite{Wagner2011} and considered
with ALRM. The overestimated real and imaginary part due to WS-HIL for
volumetric water contents above approximately 0.05 m$^3$/m$^3$ is a result of
the influence of the porosity in the mixture approach pointed out by Wagner and
Scheuermann 2009 \cite{WagScheu2009a}. With the MV-approach the relationship
between complex effective relative permittivity
$\varepsilon^\star_{r,\mbox{eff}} $ and volumetric water content $\theta$ is
predicted in the complete measured water content range.

%%-----------------------------------------------------
\section{Conclusion} The applied MV-approach gives evidence, (i) of a physical
relationship between soil dielectric relaxation behavior and soil water
characteristics as important hydraulic material property and (ii) the
applicability of multivariate methods for estimation of physico-chemical
parameters of porous media from broadband measured dielectric spectra. Against
this background, a better theoretical understanding is required of the HF-EM
material properties of porous geo-materials in general. However, this can be
only achieved if the full frequency and temperature dependence is investigated
under defined hydraulic and mechanical boundary condition \cite{Wagner2013,
Lorek2013}. Moreover, a knowledge of the frequency dependent material
properties bridge the gap between the HF-EM methods and low frequency methods
(mHz - kHz: Induced Polarization - IP / SIP, Electrical Resistivity Tomography
- ERT and kHz - MHz: Electromagnetic Methods - EM \cite{Robinson2008,
Binley2010, Revil2013, Triantafilis2013}). Therefore, from the perspective of
practical applications of broadband sensor techniques (e.g. TDR) an essential
information profit can be achieved. This, however, is inevitably linked to an
increase in the complexity of the interpretation of measurement results under
both laboratory and field conditions. Hence, the redesign of appropriate sensor
systems and probes in combination with the development of robust broadband
modeling and inversion schemes is required (c.f. \cite{Schwartz2009,
Schwartz2009a, Wagner2011, Lauer2012}).

%%-----------------------------------------------------
\section*{Acknowledgment} 
The authors grateful acknowledge the German Research
Foundation (DFG) for support of the project WA 2112/2-1 and SCHE 1604/2-1.
Moreover, the presented research is supported by a Queensland Science
Fellowship awarded to A. Scheuermann.

\bibliographystyle{IEEEtran}
\bibliography{Literatur20120516}

\end{document}